\documentclass[twocolumn]{aastex62}
\usepackage{lineno}

\usepackage[caption=false]{subfig}
\usepackage{graphicx}
\graphicspath{ {images/} }
\usepackage{hyperref}
\usepackage{amsmath}
\usepackage{booktabs}
\usepackage{xcolor}
\usepackage[percent]{overpic}
\usepackage{euler}

\begin{document}
\title{Re-examination of the time structure of the SN1987A neutrino burst data in Kamiokande-II}

\author{Yuichi Oyama}
\affiliation{High Energy Accelerator Research Organization (KEK), Tsukuba, Ibaraki 305-0801, Japan}
\affiliation{J-PARC center, Tokai, Ibaraki 319-1195, Japan}

\submitjournal{The Astrophysical Journal}
\received{October 24, 2021}\revised{November 3, 2021}\accepted{November 4, 2021}\published{December 1, 2021}

\begin{abstract}
The seven seconds' gap in the Kamiokande-II SN1987A neutrino data is reexamined. 
~~~~~~~~~~~~~~~~~~~~~~~~~~~~~~~~~~~~~~~~~~~~~~~~~~~~~~~~~~~~~~~~~~~~~~~~~~~~~~~~~~~~~~~~~~
~~~~~~~~~~~~~~~~~~~~~~~~~~~~~~~~~~~~~~~~~~~~~~~~~~~~~~~~~~~~~~~~~~~~~~~~~~~~~~~~~~~~~~~~~~
~~~~~~~~~~~~~~~~~~~~~~~~~~~~~~~~~~~~~~~~~~~~~~~~~~~~~~~~~~~~~~~~~~~~~~~~~~~~~~~~~~~~~~~~~~
\end{abstract}
\section{SN1987A neutrino burst data}
The Kamiokande-II collaboration successfully observed the neutrino burst from SN1987A,
detecting eleven neutrino events over $\sim$13~s \citep{SN1987A}.
This observation was also confirmed by the IMB collaboration \citep{IMB}.

\begin{center}
\begin{figure}[b!]
\includegraphics[width=19.5pc]{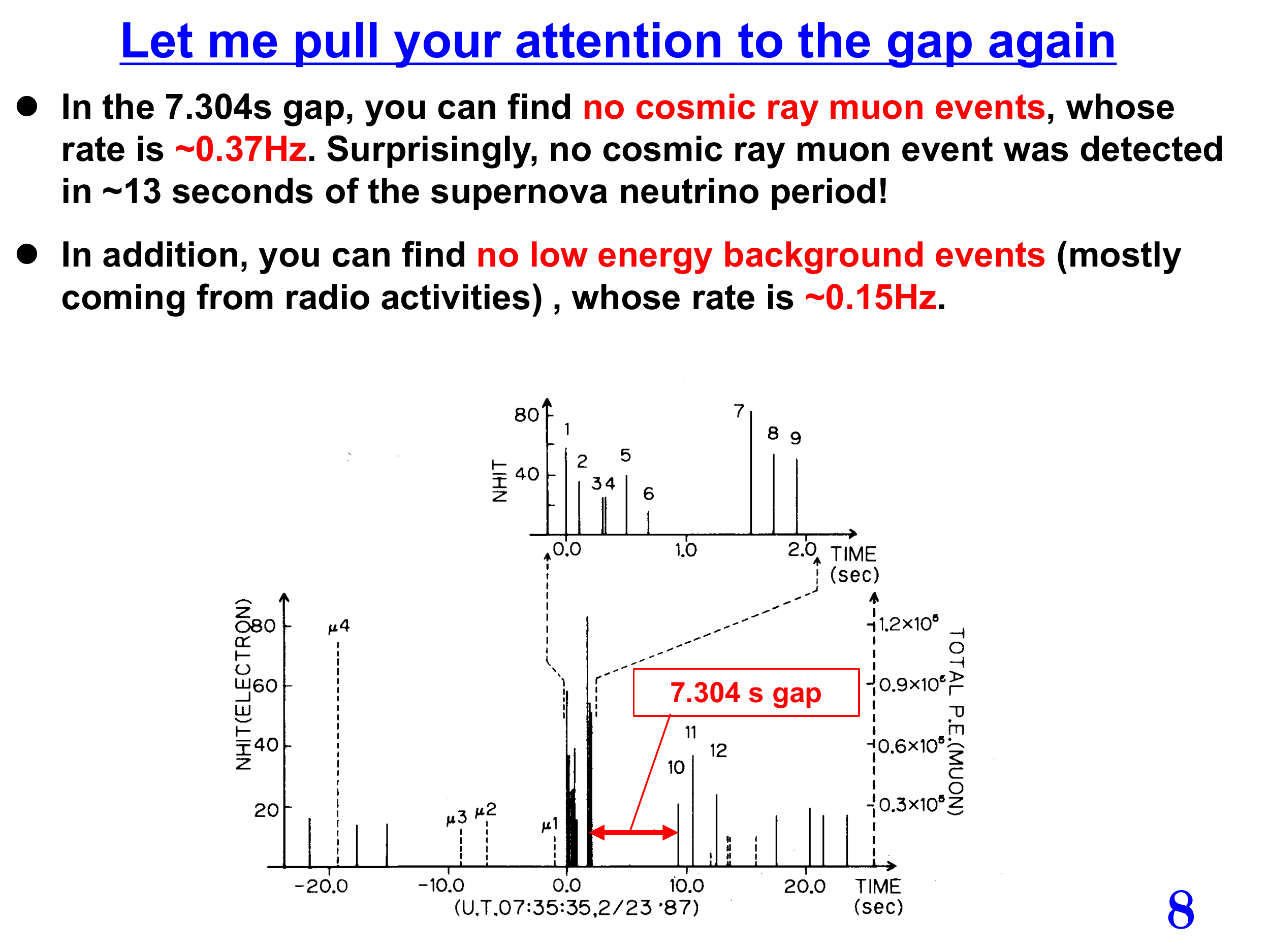}
\caption{\label{fig:SNgap}
Time structure of the Kamiokande-II events in the SN1987A period.
The original plot is shown in FIG.2 of \citet{SN1987A}.
The time period of the seven seconds' gap is also shown.
}
\end{figure}
\end{center}

The Kamiokande-II SN1987A events have a curious
time structure, known as the ``seven seconds' gap''.
Between the 9th and 10th events, there was a gap of 7.304~s,
as shown in Fig.~\ref{fig:SNgap}.
The IMB collaboration found three neutrino events in that period, and
the time gap was not due to the absence of the neutrino flux \citep{IMB,SN1987Afull}.
Assuming a uniform neutrino flux, the probability of observing a seven seconds'
gap is less than 5\% \citep{Hirata}.
However, no reasonable explanation for the gap, other than a chance
coincidence, has been found.

This study points out that no other categories of events were detected
during the 7.304~s gap.
As shown in Fig.~\ref{fig:SNgap}, no cosmic ray muon event with a rate of 0.37~Hz was detected.
Interestingly, no cosmic ray muon event was detected in $\sim$13~s of the entire SN1987A neutrino period.
Furthermore, no low-energy background event, which mostly originates from radioactivity with a rate of 0.15~Hz,
was detected.

The probabilities of detecting no SN1987A neutrino, no cosmic ray muon and
no low energy background in $\Delta t=7.304~\rm{s}$ are listed
in Table~\ref{tab:zeroevent}.  From these, 
the overall probability ($P$) that none of these events was detected can be calculated as
$$P < 0.05 \times 0.067 \times 0.334 = 0.00112.$$
A value as low as $P < 0.112\%$ is unlikely to be due to a chance coincidence.
Therefore, it is natural to consider a possible dead time of the Kamiokande-II detector.

\begin{table*}[t!] 
\caption{
Calculation of the probabilities that no event was
detected in the 7.304~s time period.
The probability for SN1987A neutrinos was obtained from
\citet{Hirata}. Other probabilities were
obtained using the Poisson probability function
${\rm Poi}(N|\lambda)={{e^{-\lambda} \lambda^{N}}\over{N!}}$.
}
\begin{center}
\begin{tabular}{lcccc}
\hline
\hline
                            &Rate& Expected \# &calculation& Probability \\
                            & (Hz) & in 7.304~s &  & of 0 event \\
\hline
SN1987A $\nu$ &&&K.S.Hirata&  $<$ 0.05\\ 
cosmic ray $\mu$  &0.37&2.70&Poi(0$|$2.70)&0.067\\ 
low energy B.G.   &0.15&1.10&Poi(0$|$1.10)& 0.334\\ 
\hline
\hline
\end{tabular}
\end{center}
\label{tab:zeroevent}
\end{table*}

\section{Possible dead time}

The block diagram of the Kamiokande-II data acquisition is shown in
FIG.2 of \citet{SN1987Afull}.
The original data were directly recorded onto 2400 feet, 6250 byte-per-inch
open-reel magnetic tapes (MTs) because of the small capacity of the hard disks.
When the data were recorded, ``write error” occasionally occurred. Partial rewind and retry
were automatically performed, and the recovery took $\sim$10~s.
During that time, the data buffers in the computer memory might become full, and no new data could be accepted.
In fact, the Kamiokande-II shift manual had the following instructions: To avoid frequent write error, clean
the head of the MT recorder using isopropyl alcohol when mounting MTs. 
The rate of the write error primarily depended on the cleanliness of the head of the MT recorder
and status of the MTs. 
If the memory of the author is correct, the rate of the error was on the order of once per hour.
The ``write error and retry” could explain the origin of the seven seconds' gap,
even though it cannot be verified.

In \citet{SN1987A}, it was reported that the $\bar{\nu}_{e}$ output of SN1987A is estimated to be
$8 \times 10^{52}$ ergs. If the seven seconds' gap is dead time and it is assumed that similar
number of neutrino events is expected during this dead time,
the $\bar{\nu}_{e}$ output of SN1987A should be larger by a factor $\sim$2.

To summarize, the probability of the Kamiokande-II experiment detecting ``no SN1987A neutrino”, ``no cosmic ray muon” and ``no low energy background” for 7.304~s is less than 0.112\%.
The possibility that the seven seconds' gap was seven seconds' dead time of
the Kamiokande-II data acquisition cannot be excluded.

\end{document}